\documentstyle[12pt]{article}

\def\ie{\hbox{i.e.}{}} 		
\def\eg{\hbox{e.g.}{}} 		\def\cf{\hbox{cf.}{}}
\def\etal{\hbox{et al.}{}} 	

\def\ltap{\;\raisebox{-.5ex}{\rlap{$\sim$}} \raisebox{.5ex}{$<$}\;}
\def\gtap{\;\raisebox{-.5ex}{\rlap{$\sim$}} \raisebox{.5ex}{$>$}\;}

\def\beq{\begin{equation}}
\def\eeq{\end{equation}}

\def\beqn{\begin{eqnarray}}
\def\eeqn{\end{eqnarray}}

\relax

\relax
\def\pl#1#2#3{Phys. Lett. {\bf #1} (19#2) #3}
\def\zp#1#2#3{Z. Phys. {\bf #1} (19#2) #3}
\def\prl#1#2#3{Phys. Rev. Lett. {\bf #1} (19#2) #3}

\def\prev#1#2#3{Phys. Rev. {\bf #1} (19#2) #3}
\def\np#1#2#3{Nucl. Phys. {\bf #1} (19#2) #3}

\def\ijm#1#2#3{Int. J. Mod. Phys. {\bf #1} (19#2) #3}

\relax

\begin{document}


\newcommand{\sss}{\scriptscriptstyle}
\def\r#1{\ignorespaces $^{\sss #1}$}

\newcommand{\ep}{\epsilon_{\sss c}}
\newcommand{\beps}{\epsilon_{\sss b}}
\newcommand{\rar}{\rightarrow}
\newcommand{\ve}{$\nu_{\sss e}$}
\newcommand{\epem}{$e^{\sss +}e^{\sss -}$}
\newcommand{\mz}{M_{\sss Z}}
\newcommand{\mw}{M_{\sss W}}
\newcommand{\mb}{m_{b}}
\newcommand{\mc}{m_{c}}
\newcommand{\rs}{\sqrt{s}}
\newcommand{\vb}{v_{\sss b}}
\newcommand{\ab}{a_{\sss b}}
\newcommand{\afb}{A_{\sss FB}^0(\mz)}
\newcommand{\as}{\alpha_{\sss S}}

\newcommand{\sth}{\sin\theta_{\sss w}}
\newcommand{\cth}{\cos\theta_{\sss w}}
\newcommand{\sthq}{\sin^{2}\theta_{\sss w}}
\newcommand{\cthq}{\cos^{2}\theta_{\sss w}}
\newcommand{\Z}{$Z^{\sss 0}$}

\newcommand{\xl}{x_{\sss \ell}}
\newcommand{\xn}{x_{\sss \nu}}
\newcommand{\yl}{y_{\sss \ell}}
\newcommand{\yn}{y_{\sss \nu}}
\newcommand{\mxl}{\langle x_{\sss \ell}\rangle}
\newcommand{\mxn}{\langle x_{\sss \nu}\rangle}
\newcommand{\mnx}{\langle x^{\sss N}\rangle}
\newcommand{\mnxb}{\langle x^{\sss N}_{\sss b}\rangle}
\newcommand{\mnxl}{\langle x^{\sss N}_{\sss \ell}\rangle}
\newcommand{\mux}{\langle x^{\sss 1}\rangle}
\newcommand{\muxb}{\langle x^{\sss 1}_{\sss b}\rangle}
\newcommand{\muxl}{\langle x^{\sss 1}_{\sss \ell}\rangle}

\newcommand{\me}{\langle E \rangle}
\newcommand{\mel}{\langle E_{\sss \ell}\rangle}
\newcommand{\men}{\langle E_{\sss \nu}\rangle}
\newcommand{\mels}{\langle E_{\sss \ell}'\rangle}
\newcommand{\mens}{\langle E_{\sss \nu}'\rangle}
\newcommand{\mpls}{\langle p_{\sss \ell}'\rangle}
\newcommand{\mpns}{\langle p_{\sss \nu}'\rangle}
\newcommand{\mes}{\langle E' \rangle}
\newcommand{\mps}{\langle p' \rangle}
\newcommand{\mga}{\langle \gamma \rangle}
\newcommand{\mgabe}{\langle \gamma \beta \rangle}

\newcommand{\mmm}{\frac{\mb}{f(\ep)}}

\setcounter{topnumber}{10}
\setcounter{bottomnumber}{10}
\renewcommand\topfraction{1}
\renewcommand\textfraction{0}
\renewcommand\bottomfraction{1}


\begin{titlepage}
\pagestyle{empty}
\begin{flushleft}
{\bf Preprint n.1009} \hfill Dipartimento di Fisica \ \ \ \ \ \ \ \ \ \\
March 17, 1994 \hfill Universit\`a di Roma ``La Sapienza'' \\
\hfill I.N.F.N. - Sezione di Roma \ \ \ \ \ \  \\
\end{flushleft}
\vspace*{\fill}
\begin{center}
{\Large \bf Lepton spectra and the b polarization \\
 at LEP}

\vspace*{\fill}

\begin{tabular}[t]{c}
{\large  Barbara Mele\r{\dagger}} \\
\\
{ \it I.N.F.N., Sezione di Roma, Italy \ \ and} \\
{ \it Dipartimento di Fisica, Universit\`a ``La Sapienza'',} \\
{       \it P.le Aldo Moro 2, I-00185 Rome, Italy} \\
\end{tabular}
\end{center}
\vspace*{\fill}
\begin{abstract}
{\small
\noindent
I review the present state of knowledge on lepton energy
 spectra in  the inclusive semileptonic decay of beauty hadrons \
($H_b\rar \ell \nu ...$) \ as a means
to measure $b$ polarization  effects on the \Z\ peak.
Charged-lepton as well as neutrino spectra are  considered.
}
\end{abstract}
\vspace*{\fill}
{\small
\r{\dagger}\ {\it e-mail address: MELE@ROMA1.INFN.IT}}
\end{titlepage}

\section{Introduction} \label{sec:intro}

\pagestyle{myheadings}
\markboth{\it B.~Mele / Lepton Spectra...}{\it B.~Mele /
Lepton Spectra...}

Heavy quark physics offers  a unique opportunity to explore the flavour
structure of the Standard Model. The third family of quarks is
the less known.
Its up-type quark, the top, has not yet been observed, although
we are getting more and more stringent limits on its mass
from direct search and radiative correction analysis.
 On the other hand, the corresponding down-type quark, the $b$, is
being studied with increasing accuracy. Both
growing statistics and the operation of dedicated vertex detector
for $b$ tagging at high energy machines already
allows a good  determination of $b$ couplings.
Present data clearly point to the $b$ as the $I_3=$--1/2 component of
a weak left-handed isodoublet \cite{kun}.

At LEP, one measures the $b$ neutral current couplings through
the $Z \rar b\bar{b}$ width  and forward-backward asymmetry
on the \Z\ peak. In fact, one has
\beq
\Gamma(Z \rar b\bar{b}) \propto (v_b^2+a_b^2) \;\;\;\;\;\;\;\;
 A_{\sss FB}^0(\mz) = \frac{3}{4} A_e A_b
\eeq
where
\beq
A_f\equiv \frac{2v_fa_f}{v_f^2+a_f^2}\;\;\;\;\;\;\;  (f=e,b)
\eeq
and I  neglected terms of the order $\beps\equiv\mb^2/\mz^2$ \cite{akv}.

In principle, another determination of the quantity $A_b$ (or,
equivalently,
of the ratio $\vb/\ab$) can be derived from the measurement of the
$b$-polarization asymmetry on the \Z\ peak.
According to the Standard Model, $b$ quarks are produced
almost completely longitudinally polarized at $\sqrt{s}\simeq \mz$.
The angular dependence of the polarization asymmetry is given
by \cite{rei}
\beq
P_b(\cos \theta)\equiv
\frac{\sigma_{\sss R}(\cos\theta)-\sigma_{\sss L}(\cos\theta)}
{\sigma_{\sss R}(\cos\theta)+\sigma_{\sss L}(\cos\theta)}
=-\frac{A_b(1+\cos^2\theta)+2A_e\cos\theta}
{1+\cos^2\theta+2A_eA_b\cos\theta}
\label{angle}
\eeq
where $\sigma_{\sss R(L)}$ is the production cross section for
a $b$ quark with right-(left-)handed longitudinal polarization
and, as above, I neglected mass effects of the order $\beps$.
Since $A_e$ is quite smaller than $A_b$,
eq.(\ref{angle}) depends only weakly on the $b$ production angle
$\theta$. The corresponding average mean polarization is
\beq \label{pbc}
P_b\equiv\frac{\sigma_{\sss R}-\sigma_{\sss L}}
{\sigma_{\sss R}+\sigma_{\sss L}}=-A_b = -\frac{2\vb\ab}{\vb^2+\ab^2}
\eeq
QCD corrections to $P_b$ have been computed in ref.\cite{kor}.
Chirality-violating mass effects in the production cross section
for polarized $b$ quarks at one-loop are found to decrease the
tree-level polarization \ $P_b=-A_b$ \ by less than $3\%$.

In analogy to the $\tau$ polarization measurement at LEP,
measuring $P_b$ would provide an alternative method
 to determine $A_b$ that has the advantage, with  respect to the
forward-backward asymmetry $A_{\sss FB}^0$,
not to be suppressed by the small quantity $A_e$ (\cf\ eq.(1)).
Furthermore, as anticipated, $P_b$ is quite large in the Standard
Model. One has (assuming $\sthq=0.23$)
\beq
P_b=-0.936
\label{pb}
\eeq
 to be confronted with the $\tau$ polarization
value $P_{\tau}\simeq-0.16$.

Unfortunately, contrary to the $\tau$ polarization asymmetries that
have already provided a nice determination of $A_{\tau}$ and $A_e$
at LEP \cite{pre},
a measurement of $P_b$ is not as much straightforward. This is mainly
due to hadronization and fragmentation
properties of $b$ quarks that largely dilute and obscure $b$
polarization effects
in the observed final states. Moreover, there are rather
large theoretical uncertainties in the predictions of such
strong interaction effects, that make the disentangling of the
polarization $P_b$ a quite difficult task.

In this paper, I will review the present status of theoretical
understanding about the possibility of detecting some effect
of such a large value of
$P_b$ in $b$ hadron final states on the \Z\ peak.
I will concentrate
on $P_b$ sensitivity of lepton energy spectra in
the inclusive semileptonic $b$ hadron
decay. Both charged-lepton and neutrino spectra will be  considered.
Theoretical ambiguities coming from our limited
knowledge of the perturbative component of the $b$ fragmentation
function will be  especially stressed.
Then, I will present some information on $P_b$ that
can be extracted from the already available LEP and PETRA data.
Recent developments on the neutrino spectrum as a
particularly efficient tool to measure a polarization effect
at LEP will be described as well.

\section{Na\"{\i}ve lepton energy spectra}
The $b$ polarization influences in a sizable way the lepton
distributions in $b$ semileptonic decays $b\rar\ell\bar{\nu}_{\ell}c$,
where $\ell$ is an electron or a muon (see also ref.\cite{tom}).
The branching ratio for these modes is relatively
high (about 20\%).
The inclusive differential distribution for semileptonic
hadron decay have been demonstrated to be equal to the parton-model
one up to corrections of order $(\Lambda_{\sss QCD}/m_q)^2$
(that in this case amount to a few percent) \cite{cha,wis}.

Assuming a free quark, with no  effects from
hadronization and QCD corrections, the normalized decay distribution
in  the charged-lepton energy is, in the $b$ rest frame, \cite{tsa}

\beqn
\frac{1}{\Gamma}\frac{d^2\Gamma}{d\yl d(\cos\alpha_{\sss \ell})}&=&
\frac{1}{f(\ep)} \frac{\yl^2(1-\ep-\yl)^2}{(1-\yl)^2}
\{3-2\yl+\ep(\frac{3-\yl}{1-\yl})  \nonumber  \\
&+& P \cos\alpha_{\sss \ell}
[1-2\yl-\ep(\frac{1+\yl}{1-\yl})]\}     \label{CM}
\eeqn
where
\beq
f(\ep)=1-8\ep+8\ep^3-\ep^4-12\ep^2 \log\ep .    \label{fep}
\eeq
In eq.~(\ref{CM}), $\yl=2E_{\ell}'/m_b$ \ ($0\leq \yl \leq 1-\ep$),
\ $E_{\ell}'$ is the $e^{\sss -}$ (or $\mu^{\sss -}$) energy
in the $b$ rest frame, $\alpha_{\sss \ell}$ is the angle between the
lepton and the $b$ spin quantization axis, and $\ep=m_c^2/m_b^2$.
In the laboratory system, the $b$ quark moves with velocity
$\beta = p_b/E_b \leq \sqrt{1-(2m_b/M_Z)^2}$. If we quantize the $b$
spin along the direction of $\vec{\beta}$, the lepton-energy decay
distribution in the laboratory frame can be obtained by
\beqn \label{BOOST}
\frac{1}{\Gamma}\frac{d\Gamma}{dx} =
\int_{y^-}^{y^+} \frac{2}{\beta y}
\frac{1}{\Gamma}\frac{d^2\Gamma}{dy d(\cos\alpha)}dy ,
\eeqn
where
\beqn
y=\yl ,\; \; \; \; \; \; \; \;
x&=&\xl\equiv\frac{E_{\ell}}{E_b} ,\; \; \; \; \; \; \; \;
\cos\alpha=\cos\alpha_{\sss \ell}=\frac{1}{\beta}(\frac{2x}{y}-1) ,
\nonumber \\
y^-&=&\frac{2x}{1+\beta}   ,  \; \; \; \; \; \; \; \; \; \; \;
y^+=\min\left\{ \frac{2x}{1-\beta}, (1-\ep) \right\} , \nonumber
\eeqn
and $E_{\ell}$ is the charged-lepton energy in the laboratory system.
In order to reproduce more realistic experimental situations,
one can impose some cut $p_{\sss T}^{\sss min}$
on the minimum allowed lepton transverse
 momentum with respect to the $\vec{\beta}$ axis
by simply  multiplying the integrand in eq.(\ref{BOOST}) by
$\theta(p_{\sss T}- p_{\sss T}^{\sss min})$, where
$p_{\sss T}=\mb y \sin\alpha/2$.

In the LEP case ($\beta\gtap 0.99 \simeq 1$), one  gets from
eqs.(\ref{CM}) and (\ref{BOOST})
\beqn
\frac{1}{\Gamma}\frac{d\Gamma}{d\xl}&=&
\frac{(1-\ep-\xl)}{3f(\ep)} \{ 5+5\xl-4\xl^2+ P (1+\xl-8\xl^2) \nonumber\\
&+&\frac{\ep}{(1-\xl)^2} [-22+5\ep+(27-4\ep)\xl+
               5\ep \xl^2-5\xl^3  \nonumber \\
&+& P (-8-17\ep+(9+28\ep)\xl+(18-17\ep)\xl^2-19\xl^3)]  \nonumber \\
&+& \frac{6\ep^2}{(1-\ep-\xl)}\log(\frac{1-\xl}{\ep})
[3-\ep+ P (3+\ep)]\} ,
\eeqn
with $f(\ep)$ defined in eq.~(\ref{fep}).

In fig.~1a,
the results for $\frac{1}{\Gamma}\frac{d\Gamma}{d\xl}$ for a na\"{\i}ve
$b$ polarization $P_b$ (solid lines) are compared with the unpolarized
case (dashed lines), assuming two different values  (1.2 and 1.5~GeV)
for the $c$-quark mass and $m_b=$5~GeV.
Here, any $b$-fragmentation effect is neglected and the final
lepton-energy fraction in the laboratory frame is given by
 \ $\xl=2E_{\ell}/M_Z$.
The harder spectra correspond to $m_c=$1.2~GeV,
the softer ones to $m_c=$1.5~GeV. One can see that a negative value of
the $b$ polarization makes the $\ell$-energy spectrum harder.

\begin{table}
\begin{center}
\begin{tabular}[tbh]{|c|l|l|l|l|} \hline
$m_c$ (GeV) &$\mxl(0)$&$\mxl(P_b)$&$\mxn(0)$&$\mxn(P_b)$\\ \hline
$ 1.2 $&\ 0.309    &\ 0.344   &\ 0.272    &\ 0.188   \\ \hline
$ 1.3 $&\ 0.303    &\ 0.337   &\ 0.268    &\ 0.185   \\ \hline
$ 1.4 $&\ 0.298    &\ 0.329   &\ 0.264    &\ 0.182   \\ \hline
$ 1.5 $&\ 0.291    &\ 0.322   &\ 0.259    &\ 0.179   \\ \hline
\end{tabular}
\vskip .3cm
{\bf Table I}: Average of the electron and neutrino
energy fractions in the laboratory frame,
$\mxl$ and $\mxn$, for different choices of $P$ and $m_c$.
Fragmentation effects are not included.
\end{center}
\end{table}
\noindent  In Table I, the mean value $\mxl$ of the
distribution is shown, for both
the polarized and the unpolarized cases, at different values of $m_c$.
By going from the unpolarized case to the na\"{\i}ve maximum $b$
polarization $P_b$, an
increase of about 10\% in the mean value of the $\ell$-energy fraction
is observed (the same
is true for the $\overline{b}\rightarrow \overline{l} \nu_l \overline{c}$
case, in the limit of exact $CP$ invariance).
However, a heavier $m_c$ softens the $\xl$ spectrum in all cases.
As far as $\mxl$ is
concerned, the polarized case with $m_c=$1.5~GeV is not very different
from the unpolarized case with $m_c=$1.2~GeV (cf. Table I).
Hence, the uncertainty on $m_c$
could partially hide polarization effects,
if only $\mxl$ was considered. As seen in fig.~1a, a more detailed study
of the spectrum beyond the first moment is required to overcome
the limited knowledge of $m_c$.

QCD corrections to the leptonic distribution in eq.~(\ref{CM}) have
 been computed in refs.\cite{ali}--\cite{czaa}.
The order-$\as$ correction for unpolarized $b$'s
 differs from a constant value
only near the end point of the spectrum, so that it amounts to a
nearly constant factor \cite{ali,cor,jez}. Hence, the shape of the
lepton-energy distribution is not significantly altered after corrections
are included.
The correction to the $P$ dependent part
of the spectrum is also found to be small \cite{cza,czaa}, especially
in the case of non-vanishing $\ep$, (\ie, $\ep\sim \mc^2/\mb^2$).

I now consider the energy spectrum of neutrinos in
$b\rar\ell\bar{\nu}_{\ell}c$. Although, in principle, missing energy
spectra are more difficult to determine experimentally, the
neutrino energy distribution turns out to be much more
sensitive to $P$ than the electron and muon distributions.
One consequence is that
the $P_b$ effect on the average neutrino energy $\mxn$ is quite larger
than the ambiguity deriving from the uncertainty on $\mc$,
contrary to the charged-lepton case.

The neutrino energy spectra for the $b\rar\ell\bar{\nu}_{\ell}c$
decay can be straightforwardly obtained from the charged-lepton spectra
in the decay $c\rar\bar{\ell}\nu_{\ell}s$ \cite{cor,cza}.
Hence, eq.(\ref{CM}) is replaced by
\beqn  \label{CMnu}
\frac{1}{\Gamma}\frac{d^2\Gamma}{d\yn d(\cos\alpha_{\sss \nu})}&=&
\frac{1}{f(\ep)}\frac{6\yn^2(1-\ep-\yn)^2}{(1-\yn)}
\left(1 + P \cos\alpha_{\sss \nu} \right).
\eeqn
where $\yn=2E_{\nu}'/m_b$,  $E_{\nu}'$ is the $\nu$ energy in the
$b$ rest-frame and  $f(\ep)$ is defined in eq.~(\ref{fep}).

In analogy with the electron case, the neutrino energy distribution
in the laboratory frame can be obtained through eq.(\ref{BOOST})
and eq.(\ref{CMnu}),
by replacing $\ell$ variables ($\xl, \yl, \cos\alpha_{\sss \ell}$)
with $\nu$ variables ($\xn, \yn, \cos\alpha_{\sss \nu}$).
At LEP ($\beta\gtap0.99\simeq 1$), one obtains
\beqn
\frac{1}{\Gamma}\frac{d\Gamma}{d\xn}&=&
\frac{2(1-\ep-\xn)}{f(\ep)} \{ 1+\xn-2\xn^2+ P (-1+5\xn-4\xn^2)
\nonumber \\
&+&\ep [-5-2\ep-4\xn +  P (5+2\ep-14\xn)] \nonumber \\
&+&\frac{6\ep^2[1 + P (2\xn-1)]}{(1-\ep-\xn)}
\log(\frac{1-\xn}{\ep})\} .
\eeqn

Fig.~1b shows the neutrino spectrum in the laboratory frame for
the unpolarized case, compared to the na\"{\i}ve $b$ polarization
case ($P=P_b$), for different values of $\mc$.
Contrary to the charged-lepton spectrum, the neutrino energy
is (considerably) softer in the polarized case.
Furthermore, as anticipated, $\mc$ ambiguities are relatively small
with respect to polarization effects. For instance, for $\mc=1.2$GeV,
$\mxn$ decreases from 0.272 ($P=0$) down to 0.188
($P=P_b$), while for $\mc=1.5$GeV, $\mxn$ goes from 0.259 ($P=0$)
to 0.179 ($P=P_b$) (\cf\ Table 1). The average neutrino energy
decreases by about $30\%$ in the polarized case.
The magnitude of this effect might therefore partly compensate
the greater difficulty of neutrino measurements.

Order-$\as$ QCD corrections to the neutrino spectrum in
eq.~(\ref{CMnu}) have  been computed and found even smaller than
in the electron case \cite{cza,czaa}.

The origin of
the different sensitivity to $P$ of charged-lepton and neutrinos spectra
 can be found in the corresponding energy dependence of
angular distributions in the $b$ rest-frame
\footnote{An analogous discussion
was made in ref.(\cite{cza}) for $c$ and $b$ charged-lepton spectra.}.
Starting from eqs.(\ref{CM}) and (\ref{CMnu}), one can define the
angular asymmetries
$\sigma_{\sss \ell,\nu}(y_{\sss \ell,\nu})$ through the equations
\beq  \label{CMas}
\frac{1}{\Gamma}\frac{d^2\Gamma}{dy_{\sss \ell,\nu}
 d(\cos\alpha_{\sss \ell,\nu})}=
\frac{d\Gamma}{dy_{\sss \ell,\nu}}
\left[1 -\sigma_{\sss \ell,\nu}(y_{\sss \ell,\nu})
\cos\alpha_{\sss \ell,\nu} \right]/2 .
\eeq
Accordingly, one has
\beqn
\sigma_{\sss \ell}(y_{\sss \ell})&=&-P\frac{(1-2\yl)(1-\yl)-\ep(1+\yl)}
{(3-2\yl)(1-\yl)+\ep(3-\yl)}  \label{as1} \\
\sigma_{\sss \nu}(y_{\sss \nu})&=&-P  \label{as2}
\eeqn
While the neutrino has a maximum polarization asymmetry in all the
energy range, the $\ell$ asymmetry has a maximum
$\sigma_{\sss \ell}^{\sss MAX}=P(1-\ep)/(1+\ep)$
 at $\yn=1-\ep$ (for positive $P$) and monotonically
falls down to $-\sigma_{\sss \ell}^{\sss MAX}/3$ at $\yn=0$,
even reverting its sign. Hence, when integrating over the lepton energy
in eq.(\ref{BOOST}),
the polarization asymmetry in the softer $\ell$-energy spectrum
partly cancels the main polarization effect coming from
large $\yl$'s.
This cancellation does not take place in the case of the neutrino that,
as a consequence, is
quite more sensitive to the $b$ polarization.

\section{Hadronization and
fragmentation effects in inclusive electron and muon spectra.}

In this section, I analyse the possibility of extracting informations
on $b$ polarization from a measure of  electron and muon
spectra from the inclusive semileptonic decay of $b$ hadrons
(for LEP data see for example ref.\cite{ale}).

Hadronization effects are expected to drastically
change the final net $b$ hadron polarization.
In ref.\cite{clo}, a detailed discussion on this subject
has been presented. Its outcome is that all spin information should be
lost in the formation of both pseudoscalar and vector $B$ mesons.
 On the contrary,  $b$ baryons are expected  in general to retain
the original heavy quark polarization. In particular, in the heavy
quark limit $m_{\sss b}\rar \infty$,  initial $P_b$ should be completely
transferred to the final $\Lambda_b$'s, which are produced either
 directly or as decay products of  heavier $b$ baryons.
Assuming that about $10\%$ of the total number of $b$ hadrons gives
rise to $b$ baryons \cite{wyl}, one expects to observe in the inclusive
semileptonic spectrum an effective polarization of the order of
$P_b/10\sim -0.1$.

By taking into account also $o(1/\mb)$ effects in the baryon
mass spectrum, one finds that in the decay of excited baryons
$(\Sigma_b,\Sigma_b^*)\rar \Lambda_b \pi$  one could produce
 partially depolarized
$\Lambda_b$'s, if the mass splitting $\Delta m(\Sigma_b,\Sigma_b^*)$
is of the order of the $\Sigma_b$ resonance
width \cite{clo,fal}. As a consequence, the exact determination
of the final $\Lambda_b$ polarization will eventually give some
information on the relative probability of $b$ hadronization into
$\Sigma_b$ and $\Lambda_b$ baryons, as well.

I now discuss the main source of uncertainty
that limits the disentangling of a polarization effect in semileptonic
energy spectra, \ie\ the ambiguities in the theoretical prediction
of the $b$-fragmentation function.
In fact, the presence of the neutrino in the final state prevents
the study of forward-backward polarization asymmetries along the spin
axis, since one can not reconstruct accurately the $b$ rest-frame.
Therefore, one can only study polarization effects on energy spectra
in the laboratory frame. This requires a good knowledge
of gluon radiation  phenomena, that alter significantly the $b$ energy
in the laboratory before its hadronization and decay and, as a
consequence, modify lepton energy spectra in $b\rar \ell \bar{\nu} c$.
 By the way, perturbative QCD conserves helicities
in the limit of negligible quark masses,
and therefore ``hard'' gluon radiation should not much affect
the quark polarization before its decay.

 It is worth stressing that the same problem affects
the determination of any change in the leptonic spectrum, that may
arise, for example, from  a possible new right-handed current that
mediates $b$ decays \cite{gro,amu}.

In ref.\cite{mel}, the uncertainties related to the $b$ fragmentation
 have been extensively discussed. To this end, the moments of the
fragmentation function $D_b(x_b,Q)$ have been introduced (see
ref.\cite{mel} for relevant definitions)
\beq
\mnxb(Q)=\int x_{\sss b}^{\sss N} D_{\sss b}(x_{\sss b},Q) dx_{\sss b} ,
\; \;\; \; \; \; N=1, 2, 3, ... .
\eeq
These  moments can be expressed by products of a perturbative part,
which includes QCD evolution-equation effects, and a
non-perturbative component, that takes into account the observed
moderate departures from the perturbative evolution at scales
near $\mb$
\beq \label{MOM}
\mnxb(Q)=\mnxb_{Pert}(Q)\mnxb_{NP} .
\eeq
The perturbative component of moments in eq.(\ref{MOM}) have been
computed to the next-to-leading order in $\as$ in ref.\cite{nas}.
In Table II, the average value of $x_b$, \
$\muxb_{Pert}(Q)$, for different
choices of $\mu_0$ (the renormalization scale) and $\Lambda_5$,
when $m_b=$5~GeV and $Q=M_Z$, is shown.
Here, \  $\Lambda_5$ is $\Lambda^{\rm \overline{MS}}_5$, as
defined by the two-loop expression of $\as$ with five light flavours.
\begin{table}
\begin{center}
\begin{tabular}[tbh]{|l|l|l|l|} \hline
$\Lambda_5$ (MeV) &$\mu_0=m_b/2$&$\mu_0=m_b$&$\mu_0=2m_b$\\ \hline
$ 150 $&\ 0.757    &\ 0.788 &\ 0.813   \\ \hline
$ 250 $&\ 0.717    &\ 0.759 &\ 0.791   \\ \hline
$ 350 $&\ 0.681    &\ 0.734 &\ 0.773   \\ \hline
\end{tabular}
\vskip .3cm
{\bf Table II}: $\muxb_{Pert}(Q)$ for different choices of
$\Lambda_5$ and $\mu_0$ at $Q=M_Z$ .
\end{center}
\end{table}
\noindent The values range from 0.68 to 0.81 for $m_b/2<\mu_0<2m_b$
and $150$ MeV$<\Lambda_5<350$ MeV.

On the other hand, the  non-perturbative component $\mnxb_{NP}$
in eq.(\ref{MOM}) can not be computed.
By the way, its precise magnitude is influenced by
the particular choice  of the renormalization scale  $\mu_0$ .
For $\mu_0\sim \mb$, one expects $\muxb_{NP}\;
\sim 1-o(\Lambda_{\sss QCD}/\mb)$.
Furthermore, it has been argued in ref.\cite{bon} that ARGUS data on
charmed hadrons suggest $\muxb_{NP}$ to be the same for $b$ meson and
 $b$ baryons at the percent level.
{}From LEP data, one obtains  $\muxb_{NP}\; \sim 0.9$
by subtracting the theoretical perturbative component
from the measured $b$-fragmentation function \cite{nas}.
Hence, the non-perturbative effect in $\muxb$ should  be comparable
in magnitude to
the maximum polarization effect one can expect in the na\"{\i}ve
charged-lepton spectrum ($\delta\mxl \sim 10\%$ for a fully
polarized initial $b$ quark).

 In order to control
the non-perturbative  component as well as a part of the
rather large uncertainties
in $\mnxb_{Pert}(Q)$,  the following strategy has been suggested in
ref.\cite{mel}.
One can consider ratios of the $N$-th moments of $D_b(x_b,Q)$
at different values of $Q$. In particular, by taking $Q=$91~GeV
($\sqrt{s}$ at LEP) over $Q=$32~GeV
(that can be thought of as the mean $\sqrt{s}$ at PETRA, as far as $b$
data are concerned \cite{mat}), the   ratio $R_b^N$
can be constructed, defined as
\beq \label{RB}
R_b^N=\frac{\mnxb(91{\rm GeV})}{\mnxb(32{\rm GeV})}=
\frac{\mnxb_{Pert}(91{\rm GeV})}{\mnxb_{Pert}(32{\rm GeV})} ,
\eeq
where I have made use of eq.~(\ref{MOM}).
In $R_b^N$, low-energy non-perturbative effects drop out.
Therefore, $R_b^N$ can be evaluated by using only the
 next-to-leading order calculation for $\mnxb_{Pert}$.
Also, in $R_b^N$ the dependence on the renormalization scale
$\mu_0$ of $\mnxb_{Pert}$ turns out to cancel.
In Table III, the value of the ratio $R_b^N$ for the first three moments
is reported as a function of $\Lambda_5$.
\begin{table}
\begin{center}
\begin{tabular}[tbh]{|l|l|l|l|} \hline
$\Lambda_5$ (MeV) &$R_b^1$&$R_b^2$&$R_b^3$\\ \hline
$ 150 $&\ 0.912    &\ 0.865 &\ 0.833   \\ \hline
$ 250 $&\ 0.903    &\ 0.852 &\ 0.817   \\ \hline
$ 350 $&\ 0.896    &\ 0.841 &\ 0.804  \\ \hline
\end{tabular}
\vskip .3cm
{\bf Table III}: The ratio $R_b^N$, defined by eq.~(\ref{RB}),
for different choices of $\Lambda_5$ and $N=1, 2, 3$.
\end{center}
\end{table}
\noindent By confronting Tables II and III, one can see that the ratio of
moments tends to be quite more stable than the original moments also with
respect to  $\Lambda_5$. For instance, one has an error on  $R_b^1$
of only about $2\%$ by varying $\Lambda_5$ in the range
$150$ MeV$<\Lambda_5<350$ MeV.

At this point, the observed leptonic spectrum in the laboratory is
obtained, in the collinear approximation, by convoluting
$D_b(x_b,Q)$ with the  polarization-dependent ``na\"{\i}ve''
leptonic spectrum in eq.~(\ref{BOOST}).
Hence, one can now compute the
ratios $R_{\ell}^N$ of the $N$-th moments of the leptonic spectrum
at $\sqrt{s}=M_Z$ over the ones at PETRA energies versus
the ``effective'' $b$ longitudinal polarization $\hat{P}$,
defined as the fraction of $P_b$ that survives hadronization processes at
LEP.
In fact, $R_{\ell}^N$ is given by $R_b^N$ times the polarization-dependent
ratio of the moments, $\mnx$, of the ``na\"{\i}ve'' lepton spectra at LEP
and PETRA
\beqn \label{RL}
R_{\ell}^N(\hat{P})&=&\frac{\mnxl_{LEP}(\hat{P})}{\mnxl_{PETRA}}=
\frac{\mnxb_{Pert}(91{\rm GeV}) \mnx_{LEP}(\hat{P})}
{\mnxb_{Pert}(32{\rm GeV}) \mnx_{PETRA}}   \nonumber \\
&=&R_b^N\frac{\mnx_{LEP}(\hat{P})}{\mnx_{PETRA}} .
\eeqn
One important point is that, at PETRA energies, $b$ quarks are
produced, to a very good approximation, through
$\gamma$ exchange alone, that is with zero longitudinal polarization.
Thus the low-energy data offer a way to calibrate
on unpolarized $b$ hadron samples.

In fig.~2, the value predicted for $R_{\ell}^N$ with $N=1$ (mean-value
ratio), is shown as a function of the ``effective''
$b$ polarization $\hat{P}$.
The band corresponds to
$\Lambda_5$ varying between 150~GeV (upper curves) and
350~GeV (lower curves)  in $R_b^N$.
A change $\Delta \hat{P}=\pm 1$ in the polarization
from the unpolarized case ($\hat{P}=0$) corresponds to a $\mp 10\%$
variation in  $R_{\ell}^1$. The study of higher order moments
is expecially useful for consistency checks on the measured value of
$\hat{P}$ (\cf\ ref.\cite{mel}). The  sensitivity to the exact value of
$m_c$ is in general reduced in the ratio, with respect to the
original moments.

In fig.~2, a comparison with available experimental
data \cite{mat} is also shown.
The dashed area corresponds to the measured value
$R_b^1({\rm exp})=0.894\pm0.015$, that has been determined
assuming $P=0$ in the LEP data.
 With this assumption,
one gets $\mux_{LEP}\simeq \mux_{PETRA}$ \ (neglecting $\beta$ effects
in the convolution of $D_b(x_b,Q)$ with the na\"{\i}ve leptonic spectrum)
and, as a consequence,
$R_{\ell}^1({\rm exp})\simeq R_b^1({\rm exp})$.
Experimental data
are already accurate enough to exclude large polarization effects
and point to a value of the ``effective'' $b$ polarization centred
around 0 and compatible with the theoretical prediction of about $-0.1$.

\section{ $\Lambda_b$ samples and neutrino spectra}

Higher statistics on  $b$ events at LEP will soon allow
to study polarization effects in
a purely $\Lambda_b$ sample, that is foreseen to retain most
of the initial $b$-quark polarization, \ie\ an order of magnitude larger
than in the previous inclusive analysis.
In refs.\cite{clo,kra,man} and \cite{hio},
some studies of exclusive decays of polarized $\Lambda_b$ are presented.
Of course, the ambiguities in the knowledge of the
$b$-quark fragmentation also apply to polarization measurements
based on semileptonic $b$-baryon decays.

A relevant question is whether,
with a reasonable statistics on semileptonic $\Lambda_b$ decays
(some data are already available \cite{lam}), LEP data will
no longer need to be normalized with lower-energy data.
One could study  the ratios
$\mnxl_{\Lambda_b}(\hat{P})/\mnxl_{B}$
of the lepton moments for  the $\Lambda_b$ sample over those
for the $B$-meson sample, both measured at LEP.
In this case, one gets
\beqn \label{RL2}
R_{\ell}^N(\hat{P})_{\Lambda_b}&=&\frac{\mnxl_{\Lambda_b}(\hat{P})}
{\mnxl_{B}}
=\frac{\mnxb_{Pert}(91{\rm GeV}) \mnx_{LEP}(\hat{P})}
{\mnxb_{Pert}(91{\rm GeV}) \mnx_{LEP}(0)}   \nonumber \\
&=&\frac{\mnx_{LEP}(\hat{P})}{\mnx_{LEP}(0)} .
\eeqn
Hence, only moments of the na\"{\i}ve
charged-lepton spectra are left in the ratio
and $b$-fragmentation effects are mostly cancelled.
The main problem with this method is that one should actually  calibrate
on the spectrum from unpolarized $\Lambda_b$'s (and not $B$'s).
Without such a normalization, a possible observed difference between
the $\Lambda_b$ and the inclusive spectra could be blamed on the model
dependent features of the exclusive $\Lambda_b$ decay.
The order of magnitude of these effects have been studied in
ref.\cite{mel} and found to be comparable to polarization effects
(\cf\ fig.~3).
Hence, a reliable model for $\Lambda_b$ semileptonic
decays is  needed before polarization effects can really be
established in this way.

Recently, a very interesting and effective method
to get rid of fragmentation effects in the $\Lambda_b$ semileptonic
sample was suggested in ref.\cite{bon}.
A new variable
\beq
y=\frac{\mel}{\men}
\eeq
given by the ratio of the mean electron and neutrino
laboratory energies in the inclusive semileptonic $\Lambda_b$
decay, has been introduced. This variable depends mainly on the $b$
rest-frame lepton angular distributions, while the dependence on the $b$
boost in the laboratory (and, hence, on $b$ fragmentation effects)
drops out.
Indeed, since at LEP $\beta\simeq 1$ , the mean lepton energy in the
laboratory is given by
\beq
\me = \mga \mes + \mgabe \mps \simeq \mga(\mes + \mps)
\eeq
where $p'$ is the longitudinal momentum in the $b$ direction of flight,
$\gamma=(1-\beta^2)^{-1/2}$,
and primed quantities refer to the $b$ rest-frame.
As a consequence, one has
\beq
y = \frac{\mel}{\men}\simeq \frac{\mels+\mpls}{\mens+\mpns}
\eeq
The average energies and longitudinal momenta in the $b$ rest-frame
can be easily computed by means of eqs.(\ref{CM}) and (\ref{CMnu}).
One finds \cite{bon}
\beqn
\mels&=&\mmm \left(\frac{7}{20}-\frac{15\ep}{4}-6\ep^2+10\ep^3
 -\frac{3\ep^4}{4}+\frac{3\ep^5}{20}-9\ep^2\log\ep
 -3\ep^3\log\ep\right)  \nonumber  \\
\mpls&=&- P \mmm \left(\frac{1}{20}-\frac{3\ep}{4}-4\ep^2+4\ep^3
 +\frac{3\ep^4}{4}-\frac{\ep^5}{20}-3\ep^2\log\ep
  -3\ep^3\log\ep \right) \nonumber \\
\mens&=&\mmm \left(\frac{3}{10}-3\ep-2\ep^2+6\ep^3
 -\frac{3\ep^4}{2}+\frac{\ep^5}{5}-6\ep^2\log\ep \right) \nonumber \\
\mpns&=&- P \mmm \left(-\frac{1}{10}+\ep+\frac{2\ep^2}{3}-2\ep^3
 +\frac{\ep^4}{2}-\frac{\ep^5}{15}+2\ep^2\log\ep \right) \nonumber
\eeqn

While the average energies are independent of polarization,
the average longitudinal momenta differ from 0 only for a polarized
quark.

In fig.~4,  $y$ is plotted versus $P$.
Besides being free of fragmentation ambiguities, $y$ has the advantage
to be much more sensitive than $\mel$ to the $b$ polarization
(\ie, for $\ep=0$, $y$ goes from 7/6 at $P=0$
up to 2 at $P=-1$). This is due to the opposite behaviour
of $\mxn$ with respect to $\mxl$ when varying $P$,
as well as to the larger sensitivity to $P$ of neutrino spectra
(\cf\ Table I). Also, the $y$ sensitivity to $\mc$ is rather mild.

Of course, neutrino spectra can be measured with less accuracy than
electron and muon spectra. Indeed, a missing energy has to be
determined in each event, whose experimental error ($\sim 3$GeV)
is limited essentially by
the resolution on the accompanying hadronic jets \cite{bon}.
This reflects in a $y$ resolution estimated as
\beq
\sigma_y \sim \frac{0.4}{\sqrt{N}}
\eeq
(with $N$ number of measured events) corresponding to an error
of about $10\%$ on $y$ with present statistics. This implies
an  accuracy of about $10\%$ on $P$ (\cf\ fig.~4).

In ref.\cite{bon}, also other variables are considered , that have
the advantage to eliminate most of  systematic errors. These are
the ratio $R_y$ and difference $D_y$ of the $y$'s corresponding to
the baryon and meson samples
\beqn
R_y&\equiv& \frac{y_{\sss baryon}}{y_{\sss meson}}= 1-K_{\sss R} P
\nonumber \\
D_y&\equiv&y_{\sss baryon}-y_{\sss meson} = -K_{\sss D} P \nonumber \\
\eeqn
where, for $0.06\ltap\ep \ltap 0.14$, one can predict
\beq
K_{\sss R}=0.66\pm0.02 \; \; \; \; K_{\sss D}=0.75\pm0.03
\eeq
Neither QCD corrections nor the adopted kinematical approximations
do contribute to $\delta K_{\sss R}$ and $\delta K_{\sss D}$ more
than 0.02.
Hence, any measured deviation of $R_y$ and $D_y$ from 1 and 0
respectively
will give a polarization measurement with very good efficiency.

\section{Conclusions}
In this paper, I have reviewed different strategies for
detecting $b$ polarization effects through lepton energy spectra
in the inclusive semileptonic decays of $b$ hadrons.
I have shown that, if one concentrate on electron and muon spectra
in the inclusive  $b$ hadrons decay, one has to deal with large
theoretical uncertainties coming from the only-partial
knowledge of the $b$ fragmentation function.
A strategy to control these ambiguities that makes use of both LEP
and lower-energy PETRA data has been outlined. By comparing
theoretical predictions with experimental data, one can already exclude
large polarization effects in the LEP inclusive leptonic spectrum.
Much more can be done with large statistics on the $b$ baryon sample,
by restricting  the polarization study  to $b$ baryons.
In fact, the corresponding $P$ effect should be about an order
of magnitude
larger than in the inclusive hadronic spectrum, since (contrary to $b$
baryons) $b$ meson, which  make up most of the hadronic sample,
 are expected not to retain any original spin information.
In principle, when restricting to $\Lambda_b$ samples,
 a realistic modelling of exclusive decays is required.
It has been shown that, if one can study also neutrino spectra with good
accuracy, the new variable  $y=\mel/\men$ is almost free of theoretical
uncertainties and optimizes the sensitivity to $P$, due also
to the very good sensitivity of neutrino spectra.

The combined use of $\ell^{\sss \pm}$ and $\nu$ spectra to measure
$b$ polarization effects certainly deserves future experimental
consideration and analysis.

\newpage
\noindent
{\Large \bf Figure captions}\\

\noindent
\underline{Fig. 1} \\
a) Charged-lepton energy spectrum in the
laboratory frame  for a na\"{\i}ve
$b$ polarization $P_b$ (solid lines) compared to the unpolarized case
$P=0$ (dashed lines), assuming different values for $\mc$. \
$b$-fragmentation effects are not included  \ ($x=2E_{\ell}/M_Z$).
 \\ b)  Same as in fig.~1a for the neutrino energy spectrum. \\

\noindent
\underline{Fig. 2} \
Prediction for $R_{\ell}^N$ with $N=1$ (average-value
ratio) versus the ``effective'' $b$ polarization $\hat{P}$.
The band corresponds to
$\Lambda_5$ varying between 150~GeV (upper curves) and
350~GeV (lower curves)  in the prediction of $R_b^N$.
The dashed area derives from the measured value of $R_b^1$ \
$R_b^1({\rm exp})=0.894\pm0.015$. \\

\noindent
\underline{Fig. 3} \
Mimicking model-dependent effects by varying the effective quark masses.
The $\ell$ energy spectra in the laboratory frame  for a na\"{\i}ve
$b$ polarization $P_b$ (solid lines) are compared with  the unpolarized
case $P=0$ (dashed lines), assuming four different sets of values for
$(m_b,m_c)$ :

\noindent 1) $(m_b,m_c)=(m_{\Lambda_b},m_{\Lambda_c})$=
(5.64~GeV,2.28~GeV);

\noindent 2) $(m_b,m_c)=(m_{B},m_{D^*})$=(5.28~GeV,2.01~GeV);

\noindent 3) $(m_b,m_c)=(m_{B},m_{D})$=(5.28~GeV,1.86~GeV);

\noindent 4) $(m_b,m_c)$=(5~GeV,1.5~GeV)  (as in fig.~1a).

\noindent $b$-fragmentation effects are not included
 \ ($x=2E_{\ell}/M_Z$). \\

\noindent
\underline{Fig. 4} \
Ratio of $\mel$ and $\men$ versus $b$ polarization for different
values of $\mc$.

\end{document}